\documentclass[prd,twocolumn,nofootinbib]{revtex4-1}
\usepackage{pifont}
\usepackage{graphicx}
\usepackage{amssymb}
\usepackage{dcolumn}
\usepackage{amsmath}
\usepackage{bm}
\usepackage{textcomp}
\usepackage{epstopdf}
\usepackage{color}
\usepackage{ulem}
\usepackage[toc,page,title,titletoc,header]{appendix}
\usepackage[colorlinks,
           linkcolor=blue,
           anchorcolor=blue,
           citecolor=blue
           ]{hyperref}
\usepackage{txfonts}
\usepackage{float}
\usepackage{multirow}
\begin{document}
\title{Nonuniform-temperature effects on the phase transition in an Ising-like model}

\begin{abstract}

In this study, we investigate the spatially nonuniform-temperature effects on the QCD chiral phase transition in the heavy-ion collisions. Since the QCD effective theory and the Ising model belong to the same universality class, we start our discussion by mimicking the QCD effective potential with an Ising-like effective potential. In contrast to the dynamical slowing down effects which delays the phase transition from quark-gluon-plasma to hadron gas, the spatially nonuniform-temperature effects show a possibility to lift the phase transition temperature. Besides, both the fluctuations and the correlation length are enhanced in the phase transition region. Furthermore, the critical phenomena is strongly suppressed like as the critical slowing down effects. The underlying mechanism is the nonzero-momentum mode fluctuations of the order parameter induced by the nonuniform temperature. Our study provides a method to evaluate the nonuniform-temperature effects, and illustrate its potential influence on analyzing the QCD phase transition signals at RHIC.

\end{abstract}

\author{Jun-Hui Zheng}
\affiliation{Center for Quantum Spintronics, Department of Physics, Norwegian University of Science and Technology, NO-7491 Trondheim, Norway}
\author{Lijia Jiang}
\email{lijia.jiang24@gmail.com}
\affiliation{Institute of Modern Physics, Northwest University, 710069 Xi'an, China}
\maketitle

\section{Introduction}
Exploring the QCD phase boundary and the critical point (CP) is one of the main goals at the Relativistic Heavy Ion Collider (RHIC) \cite{STAR2010,starnote,Busza2018,STAR20202S,STAR2021}. In the collider, a fireball forms quickly and then cools down. The QCD matter inside undergoes a phase transition from quark-gluon-plasma (QGP) to the hadronic phase. These two phases are separated by a dynamical phase transition surface in the fireball \cite{letessier_rafelski_2002,Aoki2006sf,Bazavov2012el,FUKUSHIMA201399,BZDAK20201}. Outside the surface, the hadrons and resonances scatter with each other and part of them decay. The inelastic collision between the hadronic matter finally ceases at a hypersurface named chemical freeze-out surface which is nested outside the dynamical phase transition surface \cite{Adamczyk2017}. The main experimental measurement related to the phase transition signals are event-by-event fluctuations of chemical freeze-out particle multiplicities \cite{STAR2021}. Searching the phase boundary and the CP from the dynamical process at RHIC, we have to face two basic questions. Does the dynamical phase transition boundary coincide with the equilibrium phase transition boundary in the QCD phase diagram? Are the critical behaviors kept to identify the CP?

\begin{figure}
\centering
\includegraphics[width=3 in]{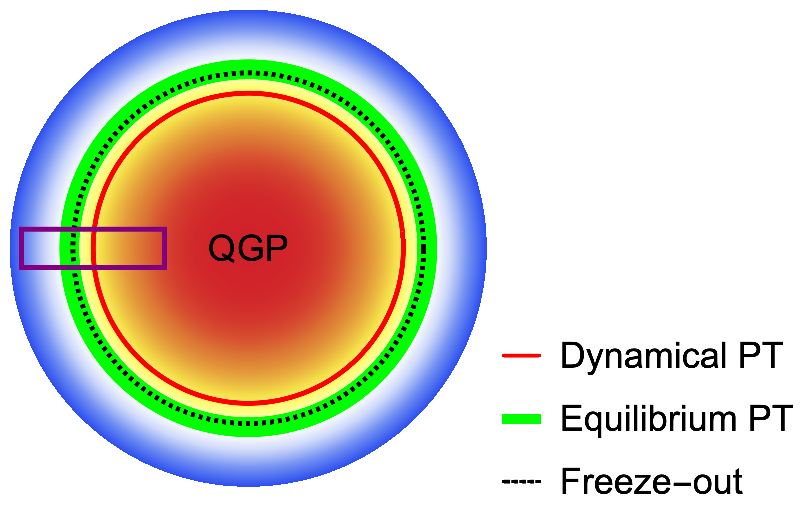}
\caption{A sketch of an instantaneous fireball. The temperature decreases from inner to outer (red to blue). The black dashed line is the chemical freeze-out surface. The green brush line refers to the isothemal surface of the equilibrium phase transition (PT) temperature in temperature-uniform systems. These two lines overlap with each other according to the lattice results and experimental data. The red solid circle represents the dynamical phase transition surface at higher temperature.}
\label{fig:fireball}
\end{figure}

Recent studies show that the chemical freeze-out line fitted from experimental data overlaps with the equilibrium phase transition boundary depicted by lattice calculation \cite{Alice2013prc,Alice2013prl,Andronic2018,Luo2020,Fukushima:2020yzx,Bazavov2014,Kaczmarek:2011zz,Bazavov:2018mes,Fu:2021oaw}. It strongly hints that the dynamical phase transition inside the fireball may happen at a temperature above the equilibrium phase transition temperature so that the hadrons have enough time to freeze out (see the sketch of an instantaneous fireball in Fig.\,\ref{fig:fireball}). This cannot be predicted by the dynamical delay effects, where the dynamical phase transition follows and memorizes the behaviors of equilibrium phase transition \cite{Berdnikov2000,Mukherjee2015,Swagato2016,Jiangvr2017,SJWu2019}. On the other side, the fluctuations and the correlation length of the QCD order parameter (i.e., the $\sigma$ field) have been broadly applied in calculating the fluctuation behaviors of observables such as net charge, baryon number and particle ratios \cite{Stephanov2009,Jeon2004,STAR20202S,STAR2021}. The correlation length has been estimated to be about $3$ fm near the CP by including the finite size effect and the critical slowing down effect \cite{Berdnikov2000,Stephanov1999}. Yet how the spatially nonuniform temperature affects the QCD phase transition at RHIC such as the phase transition point, the fluctuations, and the correlation length remains unclear.

In this paper, we investigate the spatially nonuniform-temperature effects on the QCD phase transition in a fireball context. Note that both the position and the shape of the dynamical phase transition surface vary with time during the fireball evolution. As shown in Fig.\,\ref{fig:fireball}, we take an instantaneous slender brick cell in the fireball with phase boundary located in the middle. In the brick cell, the temperature is spatially nonuniform. The phase transition region where the dynamical slowing effect is magnified is just a narrow part of the brick cell, therefore, we simplify our discussion by further supposing the relaxation of the $\sigma$ field configurations in the whole brick cell approaches to zero. As a result, the $\sigma$ field reaches its stationary distribution instantly. With this Markov assumption, the instantaneous dynamical phase transition surface turns into the stationary phase transition surface in a steady temperature-nonuniform system and no dynamical slowing effects are taken into account.

For the brick cell, we calculate the stationary solution of the $\sigma$ field, deduce and discuss the corresponding fluctuation strength and the correlation length by mapping the QCD effective potential to the Ising model. Remarkably, we find the phase transition temperature in such a temperature-nonuniform system is above the equilibrium phase transition temperature $T_c$ of a temperature-uniform system. It means if the nonuniform-temperature effects are dominant at RHIC, hadrons may form at temperature higher than the phase transition temperature determined by lattice calculations.  Further, the fluctuations and the correlation length in the phase transition region is significantly increased compared to that in the periphery of the cell, making a signal of QCD phase transition. However, the CP cannot be identified from 
the phase transition scenarios due to the nonzero-momentum mode fluctuations of the $\sigma$ field induced by nonuniform temperature. 

The rest of the paper is organized as follows. In Sec.\,\ref{tpsect}, we introduce a tanh-type nonuniform temperature profile to the brick cell, with finite temperature gradient in the phase transition region. In Sec.\,\ref{pfsect}, the probability distribution function of the order parameter field in the temperature-nonuniform system is developed. The Ising-like QCD effective potential is employed in the probability distribution function. In Sec.\,\ref{oppsect}, the stablest order parameter profile with maximum probability  is evaluated.  In Sec.\,\ref{flusect} and Sec.\,\ref{clsect}, the fluctuations around the stablest profile and the correlation length of the order parameter are calculated and analyzed, respectively. In Sec.\,\ref{exam}, we show results with a more realistic temperature profile. In Sec.\ref{sdsect}, we summarize our main results and give further discussions and outlook.

\section{Temperature profile}\label{tpsect}
First, we start the discussion by formalizing the temperature profile in the brick cell as shown in Fig.\,\ref{fig:fireball}. For simplicity, we suppose the y-z plane (the cross-section of the brick cell) is isothermal, and the temperature function along the $x$-axis (the longitude direction of the brick cell) is spatially dependent,
\begin{equation} \label{tempfun}
T(x) = T_c + \frac{\delta T}{2} \tanh\left(\frac{x}{w}\right),
\end{equation}
where $\delta T$ is the temperature bias between the two ends of the cell. The width $w$ refers to the range of the region near the equilibrium phase transition surface ($x=0$) where a finite temperature gradient ($\sim \delta T/ 2w$) presents.

Note that the real temperature profile is determined by the background matter fields like as quarks and gluons \cite{Heinz2001}.   An example with a more realistic temperature profile is presented in Sec.\ref{exam}, the results  qualitatively agree with that from the tanh-type temperature profile. 
Nevertheless, the dynamical phase transition at RHIC is complicated, for example, the baryon chemical potential profile is also spatial nonuniform. In this study, we adopt the simplified temperature profile to focus our attention on the nonuniform-temperature effects in the phase transition region. In addition, we assume the baryon chemical potential is homogeneous in the cell.

In our numerical simulation, the temperature bias for the temperature profile is set as $\delta T =40 $ MeV and the width is set to be $w=1$ fm  (or $w=0.5$ fm). The corresponding temperature gradient in the phase transition region is about $20$ MeV/fm ($40$ MeV/fm), which is comparable to the gradient in a real fireball. For example, in a fireball of radius $10$ fm with  a central temperature $200$ MeV, the mean temperature gradient along the radial direction is $20$ MeV/fm.

\section{Partition function}\label{pfsect}
As the local equilibrium assumption is proved to be well-performed in the relativistic hydrodynamics \cite{Heinz2001,Luzum2008,Gale2013,Song2007,DU2020hy,Shen2020fi}, we carry on this assumption in our calculation. Thus, the probability distribution function of the $\sigma$ field in the temperature-nonuniform system is a product of the local probability distribution function \cite{Stephanov2009} at different position $\bm r$. In the continuous limit, the probability distribution function is
\begin{equation} \label{weight}
P[\sigma(\bm r)] \propto \exp \left\{-\int d\bm r \frac{(\nabla \sigma)^2/2 + V[\sigma(\bm r)] }{T(x)}\right\},
\end{equation}
with $\bm r =(x,y,z)$.
The effective potential of the $\sigma$ field can be obtained from different QCD-inspired models \cite{Skokov2010sj,Bazavov2014,ROBERTS1994477,Jungnickel:1995fp,Schaefer:2006sr,qin2011,Jiang2013,Fukushima_2010,SCHAEFER2005479,Scavenius2001,Paech2003}. For instance, in the linear sigma model coupled to constituent quarks, the effective potential (the grand canonical potential) of the $\sigma$ field is obtained by integrating out quarks \cite{Scavenius2001}. Generally, the QCD effective potential in the CP regime can be Taylor expanded, $V[\sigma] = \sum_n z_n (\sigma-\sigma_0)^n$, where $\sigma_0$ is the minimum point of the $\sigma$ field at CP $(\mu_c, T_c)$. Since the QCD effective theory and the Ising model belong to the same universality class, we assume that the effective potential can be parameterized as $V[\sigma] = h (\sigma-\sigma_0) + r (\sigma-\sigma_0)^2 + c (\sigma-\sigma_0)^4$, where $r$ is the reduced temperature and $h$ is the magnetic field in the Ising model \cite{Jiang2017,Stephanov2019}. In the simplest linear mapping between $(T,\mu)$ and the Ising variables $(h,r)$ \cite{Stephanov2019, Nonaka2005,Stephanov2011,Jiang2017}, we have $h =  a \Delta T$  and $r = b 
\Delta \mu $, where $\Delta T = T - T_c$ and $\Delta \mu = \mu-\mu_c$. Consequently, we obtain
\begin{equation}\label{isingmodel}
V[\sigma] = a(T - T_c)(\sigma-\sigma_0) + b (\mu-\mu_c) (\sigma-\sigma_0)^2 + c (\sigma-\sigma_0)^4, 
\end{equation}
where $a>0$, $b<0$ and $c>0$ are free parameters which can be constrained by the QCD effective theories, lattice calculations or experimental data etc. Note that in a general linear mapping, the linear transformation between $(h,r)$ and $(\Delta T, \Delta \mu)$ contains two mixing angles \cite{Stephanov2019}. We omit these angles in this article for simplicity. Within the simplest mapping, the phase transition temperature is $\mu$ independent. For $\Delta \mu \equiv \mu-\mu_c>0$ and $\Delta\mu <0$, the effective potential describes the first-order phase transition and crossover respectively as the change of temperature.

Throughout the article, we set $a= 0.5$ fm$^{-2}$,  $b=-0.25$ fm$^{-1}$, and $c=3.6$. These values are chosen by constraining the correlation length and the expectation values of the $\sigma$ field in the reasonable ranges as is explained below. The phase transition temperature is set to $T_c =160 $ MeV, which is close to the lattice simulation result \cite{Bazavov2014,Kaczmarek:2011zz,SJWu2019}. In this parameter setting, for $\Delta \mu =0$, $\Delta T = \pm 20 $ MeV, the minimum point of the $\sigma$ field (i.e. the expectation value in the mean-field approximation) is $\sigma  = \sigma_0 \mp (a \Delta T/ 4c)^{1/3}= \sigma_0 \mp 30 $ MeV and the correlation length of the $\sigma$ field is about 1 fm (which is a natural value of correlation length away from the CP \cite{Berdnikov2000, Jiang2016}). For $\Delta T =0$, $\Delta \mu = 200 $ MeV, the expectation value is $\sigma  = \sigma_0 \pm \sqrt{-b\Delta\mu/2c}  = \sigma_0 \pm 37 $ MeV and the correlation length of the $\sigma$ field is 1 fm. Different choices of the value of $(a,b)$ are equivalent through rescaling the magnitude of $\Delta T$ and $\Delta \mu$. The empirical value of $\sigma_0$ at CP is around $45$ MeV \cite{Scavenius2001,Paech2003}. Since the value of $\sigma_0$ will not influence our discussion on fluctuations and correlation length, we simply set $\sigma_0=0$ in the following. Then, in thermal equilibrium, $\sigma<0$ and $\sigma>0$  correspond to the QGP phase and the hadron phase, respectively.

\section{The stablest order parameter profile}\label{oppsect}
In this section, we figure out the stablest order parameter profile which maximizes the probability. Since the temperature is spatially nonuniform, the  local order parameter which maximizes the probability distribution function is never again determined by minimizing the effective potential $\partial V[\sigma]/\partial\sigma =0$, but satisfies the extreme value condition, $\delta P[\sigma]/\delta\sigma =0$. Explicitly, we have
\begin{eqnarray} \label{dpds}
\delta P[\sigma]  &=& - P[\sigma] \int d\bm r \left[\frac{\nabla \sigma \cdot  \nabla \delta\sigma}{T(x)} + \frac{\partial V}{
\partial \sigma} \frac{\delta\sigma }{T(x)}\right] \notag\\
&=& - P[\sigma] \int d\bm r \delta\sigma \left[-\nabla  \cdot \frac{\nabla \sigma}{T(x)} + \frac{1}{T(x)}\frac{\partial V}{ 
\partial \sigma} \right].
\end{eqnarray}
The formula in the bracket vanishes in the extreme value condition $\delta P[\sigma]/\delta\sigma =0$. Therefore, we have
\begin{equation} \label{classical}
 \nabla^2 \sigma = \frac{1}{T} \nabla T \cdot \nabla \sigma + \frac{\partial V}{ \partial\sigma}.
\end{equation}
As we have supposed that the temperature distribution in the y-z plane is isothermal, the $\sigma(\bm r)$ that maximizes the weight function must be flat in this plane. Thus $\sigma(\bm r)$ depends only on $x$, and Eq.\,\eqref{classical} reduces to a one-dimensional problem. The boundary condition is given by the local order parameters at the ends, i.e., $\sigma(x=-L/2) = \sigma_L $ and $\sigma(x=L/2) =\sigma_R$, where $\sigma_L$ and $\sigma_R$ are the global minimum point of the potential $V[\sigma]$ at $x=\mp L/2$ and $L$ is the cell's length. Note that when $L$ is sufficient large, i.e., $L \gg w$, the magnitude of $L$ will not influence the following results. 

\begin{figure}[hbtp!]
\centering
\includegraphics[width=3.2 in]{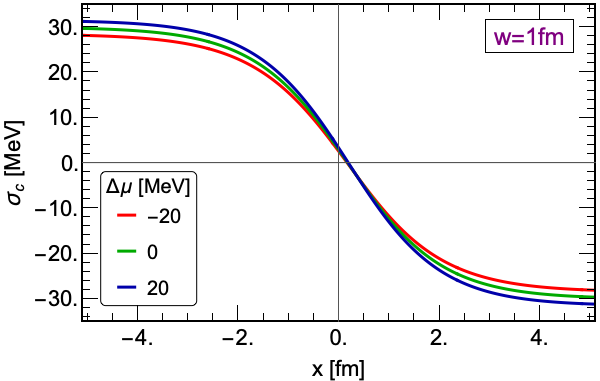}
\caption{The order parameter profile $\sigma_c(x)$ in the brick cell, with the red, green and blue lines represent results in the crossover ($\Delta\mu<0$), CP ($\Delta\mu=0$) and the first-order phase transition ($\Delta\mu>0$) scenarios, respectively. The phase transition point (which is corresponding to $\sigma_c(x)=0$) locates at some position $x_{PT}>0$.}
\label{fig:classical}
\end{figure}

The solution $\sigma_c(x)$ to  Eq.\,\eqref{classical} is presented in Fig.\,\ref{fig:classical}, with $w=1$ fm and different $\Delta\mu$. A main information from these order parameter profiles is that $\sigma_c(x)$ changes its sign at $x>0$,  no matter the sign and magnitude of $\Delta\mu$.
It is easy to check that, without the temperature gradient term $(1/T)\nabla T\cdot \nabla \sigma$,  the solution $\sigma_c(x)$ is an odd function of $x$ and vanishes at $x=0$. As the  $(1/T)\nabla T\cdot \nabla \sigma$ term is always negative ($\partial_x \sigma <0$ and $\partial_x T>0$), it will always contribute similar corrections to the solution $\sigma_c(x)$, and the sign change of $\sigma_c(x)$ will universally happens at $x>0$. This result can be comprehended directly from the probability distribution function Eq.\eqref{weight}. In the brick cell, the hot part with high temperature is more easily fluctuated than the cold part. Therefore, $\sigma_c(x)$  will tend to the order parameter value of the cold part, and $\sigma_c(0)$ becomes positive.

\begin{figure}[hbtp!]
\centering
\includegraphics[width=3.2 in]{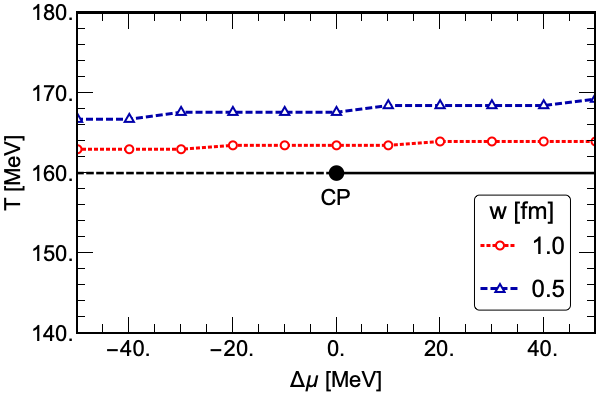}
\caption{A comparison of the phase transition temperature in  the temperature-nonuniform (red and blue dotted lines) and temperature-uniform systems (black lines). The red dotted line is with $w=1.0$ fm, and the blue dotted line is with $w=0.5$ fm.}
\label{fig:temp}
\end{figure}

Like as the equilibrium phase transition of the Ising model, we identify the point of sign change of $\sigma_c$ as the phase transition point at different $\Delta\mu$. The phase transition point always locates at some position $x_{PT} >0$ (see Fig.\,\ref{fig:classical}) and the corresponding phase transition temperature $T(x_{PT})$ is generally higher than the equilibrium phase transition temperature $T_c = T(x=0)$. Note that the phase transition temperature at the phase transition position $x_{PT}$ can be evaluated from the function of temperature profile \eqref{tempfun}. In Fig.\,\ref{fig:temp}, we show the phase transition temperature for the two widths $w=1$ fm and $w=0.5$ fm. The phase transition temperature is lifted about $3$ MeV and $8$ MeV from $T_c$, respectively. A steeper temperature gradient leads to a higher phase transition temperature.

Note that the lifted values of temperature is not universal and depend on the temperature profile.  At RHIC, the spatial temperature profile usually is not a $\tanh$-type, thus in Sec.\ref{exam}, we consider a more realistic temperature profile fitted from the hydrodynamics' output. The phase transition temperature is also lifted, which qualitatively agrees with the result from the $\tanh$-type temperature profile. We conclude that the nonuniform-temperature effects will change the phase transition temperature, and provide a possibility that the QCD phase transition happens at temperature higher than the lattice $T_c$.  In the following, we keep our discussion on the $\tanh$-type profile and reveal how the temperature profile influences the fluctuations and correlation length.

\section{Thermal fluctuations}\label{flusect}
In this section, we study the fluctuation behaviors of the $\sigma$ field and show how it is influenced by the temperature profile. We express the $\sigma$ field as the combination of the variational extremum solution and a small fluctuation, $\sigma(\bm r) = \sigma_c(x) + \delta \sigma(\bm r)$. Then, the probability distribution function $P[\sigma]$, up to the fourth order of the fluctuation, becomes
\begin{eqnarray}
P[\sigma] &\propto& \exp \left\{-\int d\bm r \frac{[\nabla ( \sigma_c + \delta \sigma(\bm r))]^2/2 + V[ \sigma_c(x) + \delta \sigma(\bm r)] }{T(x)}\right\} \notag\\
 &=&\exp \left\{-\int d\bm r \frac{[\nabla ( \sigma_c)]^2/2 +  V[ \sigma_c(x)]}{T(x)}\right\} \notag\\
&& \exp \left\{-\int d\bm r \left[\frac{\nabla \sigma_c \cdot \nabla\delta \sigma(\bm r)}{T(x)} + \frac{\delta V}{\delta \sigma}\Bigg|_{\sigma=\sigma_c} \frac{ \delta \sigma}{T(x)} \right] \right\} \notag\\
&& \exp \left\{-\int d\bm r \left[\frac{(\nabla\delta \sigma)^2}{2 T(x)} + \frac{\delta^2 V}{2(\delta \sigma)^2}\Bigg|_{\sigma=\sigma_c} \frac{ (\delta \sigma)^2}{T(x)} \right] \right\} \notag\\
&&\exp \left\{-\int d\bm r \frac{\delta^3 V}{3!(\delta \sigma)^3}\Bigg|_{\sigma=\sigma_c} \frac{ (\delta \sigma)^3}{T(x)} \right\} \notag \\
&&\exp \left\{-\int d\bm r \frac{\delta^4 V}{4!(\delta \sigma)^4}\Bigg|_{\sigma=\sigma_c} \frac{ (\delta \sigma)^4}{T(x)} \right\}. \label{flucp}
\end{eqnarray}
The first term is a finite number which depends on the profile $\sigma_c$, the second term equals to $1$ because $\delta P/\delta\sigma$ vanishes for $\sigma=\sigma_c$ [see Eqs.\,\eqref{dpds} and \eqref{classical}], and the last three terms are contributions from the fluctuations. For the Ising-like potential \eqref{isingmodel}, we have
\begin{equation}\label{wf}
P[\sigma] \propto \exp \Bigg[-\int d\bm r \frac{{(\nabla \delta\sigma)^2}/{2}  +{m^2 \delta\sigma^2  }/{2} + 4 c \sigma_c \delta\sigma^3 + c \delta\sigma^4}{T(x)}\Bigg],
\end{equation}
where the mass term
\begin{equation}\label{mass}
m^2(x) = 2 b \Delta\mu + 12 c \sigma_c^2
\end{equation} 
is spatially dependent. In this article, we mainly focus on the variance of the fluctuations, so we omit the cubic and quartic terms which are of higher order of $\delta \sigma$ and can be neglected in the perturbation theory \cite{Stephanov2009}. The cubic and quartic terms will be taken into account for the higher-order cumulants of the fluctuations \cite{Stephanov2009,Jiang2016,JIANG2016360}.

\begin{figure}[tp!]
\centering
\includegraphics[width=3.2 in]{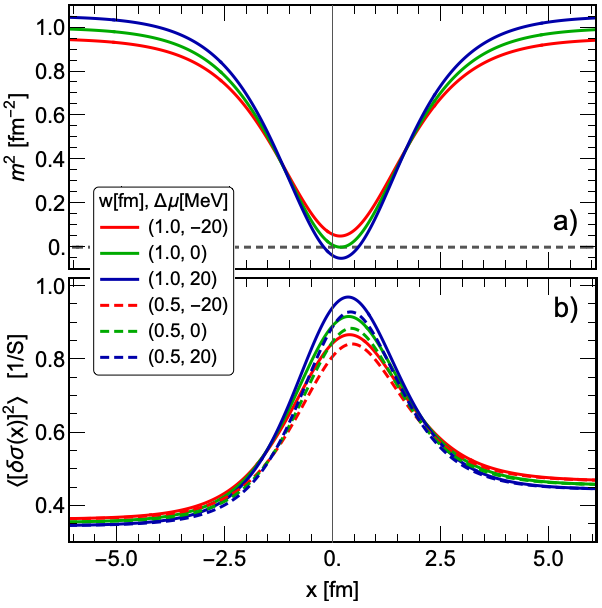}
\caption{ Panel a) presents the local mass square of the fluctuating $\sigma$ field, and panel b) presents the variance for different $w$ and $\Delta\mu$. In both panels, the red, green and blue lines represents results in the crossover ($\Delta\mu<0$), CP ($\Delta\mu=0$) and the first-order phase transition ($\Delta\mu>0$) scenarios, respectively. Solid lines are results with $w=1$ fm, and dotted lines are results with $w=0.5$ fm.}
\label{fig:mass}
\end{figure}

Conventionally, we start the discussion from the mass term of the $\delta \sigma$ field. Note that in a uniform system with temperature $T$, the correlation length is related to the mass of the $\sigma$ field: $\bar \xi = 1/\bar{m}$, where $\bar{m}^2 = \partial^2 V/\partial \sigma^2\big|_{\sigma = \bar{\sigma}} \geq 0$ and the expectation value $\bar\sigma$ is determined by the condition $\partial V/\partial \sigma\big|_{\sigma = \bar{\sigma}}=0$. Similarly, for the nonuniform case, we define a local correlation length: $\xi_{local}(x) =1/\sqrt{m^2(x)}$. We present the results of $m^2(x)$ in the brick cell in Fig.\,\ref{fig:mass}a). In the periphery, we have $m^2 \approx 1$ fm$^{-2}$ and thus  $\xi_{local} \approx 1$ fm,  which coincides with $\bar \xi$ at temperature $T=T(\pm L/2)$. This is due to the fact that the temperature becomes flat when the position is far from the center ($|x|/w \gg 1$). In the central part, $m^2(x)$ presents exotic behaviors for different phase transition scenarios. In the crossover regime ($\Delta\mu<0$), $m^2(x)>0$ everywhere. For the critical value ($\Delta\mu =0$), $m^2(x)$ vanishes at $\sigma_c =0$, and the local correlation length $\xi_{local}$ diverges. However, in the first-order phase transition regime ($\Delta \mu>0$),  $m^2(x)$ is negative in the phase transition region, which is in contrast to the positive $\bar{m}^2$ in a temperature-uniform system. Therefore, the current definition of the local correlation length is not appropriate in the phase transition region with a finite temperature gradient. As we will show below, the variance of the local fluctuation $\delta\sigma(x)$ is always positive, and is better-suited for the description of the temperature-nonuniform system.

In the following, we calculate the variance of the fluctuation. We presume the size along the $y$ and $z$ direction is much smaller than the unknown correlation length. Therefore, we can adopt the  zero-momentum mode approximation for $y$ and $z$ directions and thus $\delta\sigma(\bm r)$ depends only on $x$. The cross-section of the brick cell is denoted as $S$. Discretizing the $x$-axis with spacing length $\Delta x$, the probability distribution function becomes
\begin{equation}
P[\sigma] \propto \exp \left\{-\frac{S}{2}\sum_{i,j} \delta\sigma_{i} M_{ij}  \delta\sigma_{j}\right\},
\end{equation}
where the nonzero elements of the matrix $M$ are
\begin{eqnarray}
&&M_{ii} = \frac{1}{\Delta x}\left[\frac{1}{T_{i-1/2}} + \frac{1}{ T_{i+1/2}}  \right] +
  \frac{ m_i^2 \Delta x }{T_i}, \\
  && M_{i,i+1} =M_{i+1,i}=-\frac{1}{T_{i+1/2}\Delta x}.
\end{eqnarray}
Here, `$i$' refers to the position $x=i \Delta x$. The matrix $M$ must be positive-definite so that the solution $\sigma_c$ is guaranteed to maximize the probability distribution function. We would like to emphasize the necessity and importance of the kinetic energy in $P[\sigma]$ (see Eq.\,\eqref{wf}), which is nonzero and solves the negative $m^2(x)$ problem in the first-order phase transition scenario.  This is because in the brick cell, $m^2(x)$ constructs a potential well as shown in Fig.\,\ref{fig:mass}a), and the kinetic term has to be finite due to the uncertainty principle. From the same reason, at $\Delta\mu =0$, the fluctuations on the CP is not divergent due to a positive ground energy of $M$. The nonzero kinetic energy represents the contribution from the nonzero-momentum mode fluctuations of the $\sigma$ field, which plays a crucial role in the temperature-nonuniform system.

The variance of the local fluctuations is
\begin{equation}
\label{variance} \langle [\delta\sigma_i]^2\rangle = \frac{[M^{-1}]_{ii}}{S}.
\end{equation}
In Fig.\,\ref{fig:mass}b), we plot the results of the variance for different $w$ and $\Delta\mu$. Note that the maximum point of the variance locates a little right of the minimum point of $m^2(x)$, because the fluctuations in the right of the cell is lifted due to a higher temperature compared to the left (see Eq.\,\eqref{wf}).
Interestingly, the fluctuations on the phase transition point monotonically increase from the crossover ($\Delta \mu < 0$) to the first-order phase transition ($\Delta \mu > 0$). There are no exotic behaviors to characterize the CP ($\Delta \mu =0$). In addition, the fluctuations near the phase transition point are enhanced as the increase of the width $w$ for all the three scenarios. This can be understood in the extreme case that when $w\rightarrow \infty$, the temperature is flat locally and the fluctuations near the CP become divergent.

\begin{figure}[bhtp!]
\includegraphics[width=3.2 in]{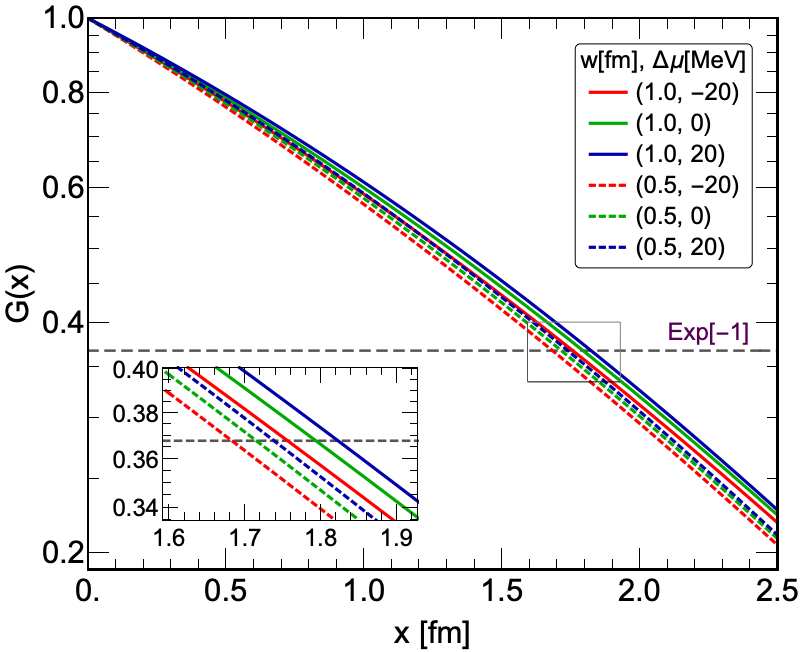}
\caption{The normalized nonlocal correlation $G(x)$ near the phase transition point in a logarithmic scale. The legends are the same as that in Fig.\ref{fig:mass}. The inset is an enlargement of the cross region marked in the plot.}
\label{fig:fluctuation}
\end{figure}

\section{Correlation length}\label{clsect}
Now we calculate the correlation length near the phase transition point from the normalized nonlocal correlation,
\begin{equation}
    G(x) = \frac{\langle \delta\sigma(x_p + {x}/{2})\delta\sigma(x_p - {x}/{2})\rangle}{\langle \delta\sigma(x_p )\delta\sigma(x_p)\rangle},
\end{equation}
where $x_p=i_0 \Delta x$ denotes the spatial location of the maximum point of the variance. Numerically, we have $G(2j\Delta x) = [M^{-1}]_{i_0-j, i_0+j}/ [M^{-1}]_{i_0, i_0}$. We plot the result in Fig.\,\ref{fig:fluctuation}. The normalized nonlocal correlation does not exactly decay  exponentially, so we determine the correlation length $\xi$ by requiring $G(\xi) =\exp(-1)$. The $\xi$ again smoothly increases from the crossover regime ($\Delta\mu<0$) to the first-order phase transition regime ($\Delta\mu>0$), and decreases as the increase of the temperature gradient. With the current parameter set, our estimation of the correlation length $\xi $ is about $1.65$fm $-$ $1.9$fm in the central part of the brick cell,  which is significantly larger than $\xi \approx 1$ fm in the periphery \footnote{Note that the results quantitatively depend on the parameter setting. With a different parameter setting, $a=0.22$ fm$^{-2}$, $b=-0.1$ fm$^{-1}$, and $c=1.6$, the correlation length is $\xi \approx 1.5$ fm in the periphery and  $\xi \approx 2.55$ fm at the phase transition point. 
The phase transition temperature is also lifted from $T_c$, and the value is about $6$ MeV for $w=1$ fm  and $11$ MeV for $w=0.5$ fm. 
The qualitative results are not changed.}. 
It's important to point out that for the critical value $\Delta \mu =0$, the correlation length does not diverge and is strongly suppressed by the nonuniform-temperature effects. The suppression is comparable to that from the critical slowing down effects \cite{Berdnikov2000}. The magnitude of the correlation length will be further suppressed when the critical slowing down effects are included.

\section{An example with a more realistic temperature profile}\label{exam}
\begin{figure}[tp!]
\centering
\includegraphics[width=3.2 in]{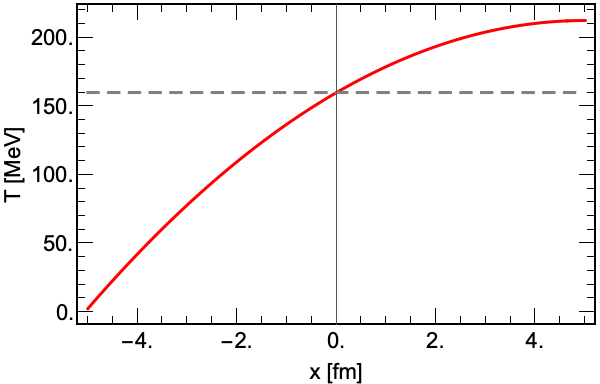}
\caption{The temperature profile in the real space. Here, for $x=0$, $T=T_c$. The position $x=5$ fm locates in the center of the fireball and the position $x=-5$ fm is the left boundary of the fireball.}
\label{fig:temprofile}
\end{figure}

\begin{figure}[tp!]
\centering
\includegraphics[width=3.2 in]{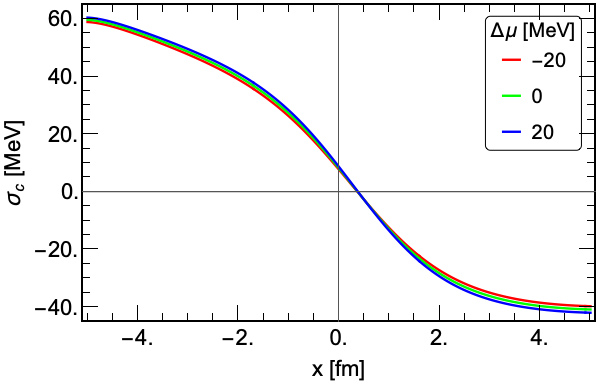}
\caption{The order parameter profile $\sigma_c(x)$, with the red, green and blue lines represent results in the crossover ($\Delta\mu<0$), CP ($\Delta\mu=0$) and the first-order phase transition ($\Delta\mu>0$) scenarios, respectively.}
\label{fig:sigmafield}
\end{figure}

In this section, we present the results with a more realistic temperature profile. The temperature profile is extracted from the hydrodynamic simulation on the fireball evolution at RHIC (after smoothening) \cite{DU2020hy}. We again set the temperature in the slender brick cell. The temperature profile is shown in Fig.\,\ref{fig:temprofile}, where the temperature at $x=0$ is the phase transition temperature $T(x=0)=T_c$, the position $x=5$ fm corresponds to the center of the fireball, and the position $x=-5$ fm represents the left boundary of the fireball. We keep all the other parameters unchanged, and further assume that the effective potential \eqref{isingmodel} is valid in the whole temperature region. For the current temperature profile, the corresponding extreme solution $\sigma_c(x)$ is shown in Fig.\ref{fig:sigmafield}. In this plot, we can find that the phase transition happens at $x_{PT}>0$, where $T\approx 168 $ MeV is about $8$ MeV larger than $T_c$. This qualitatively agrees with the result from the tanh-type temperature profile.

\begin{figure}[tp!]
\centering
\includegraphics[width=3.2 in]{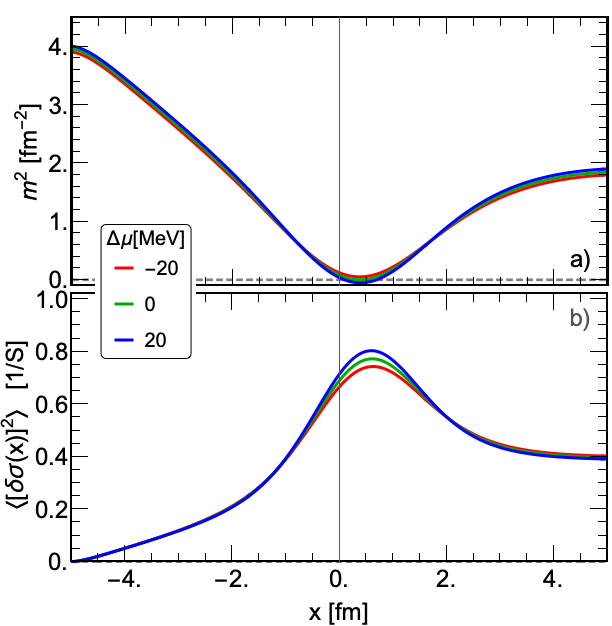}
\caption{Panel a) and b) present the local mass square and the variance of the fluctuating $\sigma$ field, respectively. In both panels, the red, green and blue lines represents results in the crossover ($\Delta\mu<0$), CP ($\Delta\mu=0$) and the first-order phase transition ($\Delta\mu>0$) scenarios, respectively.}
\label{fig:massvari}
\end{figure}

\begin{figure}[tp!]
\centering
\includegraphics[width=3.2 in]{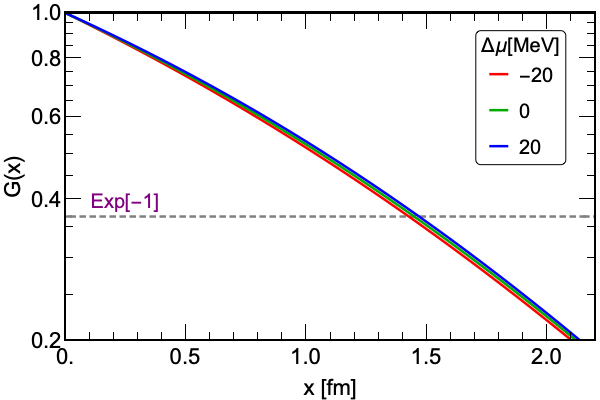}
\caption{The normalized nonlocal correlation $G(x)$ near the phase transition point in a logarithmic scale.}
\label{fig:collen}
\end{figure}

In Fig.\,\ref{fig:massvari}, we show the results of local mass square and the variance of the fluctuations. The local correlation lengths at $x=\pm 5$ fm are $\xi_{local} = 1/m\approx 0.73$ fm and $0.5$ fm, respectively. The local correlation lengths also becomes ill-defined in the first-order phase transition region, since local mass square becomes negative when $\sigma_c^2 < -b \Delta \mu/6 c$. The variance vanishes at $x=-5$ fm since $T\rightarrow 0$ at the boundary of fireball. In Fig.\,\ref{fig:collen}, the normalized nonlocal correlation near the phase transition point is plotted. The correlation length at the phase transition point is about $1.45$ fm, which is significant larger that $\xi_{local}$ at  $x=\pm 5$ fm. The variance and correlation length with this temperature profile present similar behaviors as those in the case of tanh-profile temperature.

\section{Summary and discussion}\label{sdsect}
In this article, we studied the nonuniform-temperature effects on the stablest order parameter profile, the fluctuations and the correlation length. Remarkably, we find that the phase transition temperature is generally ahead for different temperature gradients in our temperature profile settings. This hints at a possibility that if the nonuniform-temperature effects are manifest at RHIC, the hadrons may form at a temperature higher than the lattice $T_c$. In addition, the phase transition region can be identified by the enhancements of both the fluctuations and the correlation length, and the enhancements decrease as the increase of temperature gradient. However, the uniqueness of the CP behaviors are wiped off.  These novel phase transition behaviors inherit from the nonzero-momentum mode contribution of the order parameter induced by the nonuniform temperature distribution in space.

Emphasize again that as the first attempt to discuss the nonuniform-temperature effects, we keep the model and parameter settings simple to manifest the main results from nonuniform-temperature effects. The real temperature profile as well as the baryon chemical potential profile at RHIC vary for different events at different time, and they are also affected by dynamical factors like the fluctuations, jet, and flow etc. Our use of the simplest Ising mapping for the QCD potential and the assumption of both uniform chemical potential profile and tanth-type temperature profile may be oversimplified for the fireball in RHIC. On the other hand, the higher order corrections of the order parameters in the QCD effective potential is also neglected (for example, the $\sigma^5$ term related to the $h$ and $r$ mixing as discussed in Ref. \cite{Stephanov2019}). These approximations may induce uncertainty for our numerical results.

Even in the Markov approximation within our assumption, statistical average over fluctuations in different temperature profiles is needed. Different parameter setting shows that both the phase transition temperature shift and the variance of the $\sigma$ field qualitatively agree with each other. Therefore, the statistical average will not qualitatively change our conclusions within the current model setup.

In the current treatment we simplify our model to the one-dimension case by assuming the fluctuations in the cross section are frozen. For the spherical symmetry case, Eq.\eqref{classical} can also be simplified to a one-dimensional differential equation by using the spherical coordinates.
For a more complicated system related to the recent experimental data of net-proton fluctuations in Au+Au collisions \cite{STAR20202S}, a full treatment of the three-dimensional differential Eq.\eqref{classical} can be developed numerically.

As for the  phenomenological applications, we call attention to the nonuniform-temperature effects on modeling the dynamical phase transition during the fireball expansion at RHIC. The temperature gradients in the fireball is large, so the nonuniform-temperature effects should be significant but have been overlooked by now. Indeed, the nonuniform-temperature effects and the dynamical memory effects   \cite{Mukherjee2015,Herold2016prc,Jiang2017,Stephanov2018,Nahrgang2019hf,Shuryak2019prc,Rajagopal2020,Bluhm:2020mpc,Du2021} are two extreme cases corresponding to spatial correlation dominant and temporal correlation dominant, respectively. 
In our calculations, the enhancements of the fluctuations and correlations in the phase transition region show again the importance of the dynamical effects. The two effects are highly possible to interrelate with each other in the realistic fireball expansion. The combination of the nonuniform-temperature effects and the dynamical effects will provide a better description to the phase transition at RHIC. The inclusion of the nonuniform-temperature effects on the study of phase transition in the compact stars is also promising.

\begin{acknowledgements}
This work was supported by the Research Council of Norway through its Centres of Excellence funding scheme (project no.\,262633, ``QuSpin''). Lijia Jiang thanks Xiaofeng Luo and Bao-Chi Fu for helpful discussions. Jun-Hui Zheng and Lijia Jiang contribute equally to this work.
\end{acknowledgements}

\bibliographystyle{apsrev4-1}

\begin{thebibliography}{58}%
\makeatletter
\providecommand \@ifxundefined [1]{%
 \@ifx{#1\undefined}
}%
\providecommand \@ifnum [1]{%
 \ifnum #1\expandafter \@firstoftwo
 \else \expandafter \@secondoftwo
 \fi
}%
\providecommand \@ifx [1]{%
 \ifx #1\expandafter \@firstoftwo
 \else \expandafter \@secondoftwo
 \fi
}%
\providecommand \natexlab [1]{#1}%
\providecommand \enquote  [1]{``#1''}%
\providecommand \bibnamefont  [1]{#1}%
\providecommand \bibfnamefont [1]{#1}%
\providecommand \citenamefont [1]{#1}%
\providecommand \href@noop [0]{\@secondoftwo}%
\providecommand \href [0]{\begingroup \@sanitize@url \@href}%
\providecommand \@href[1]{\@@startlink{#1}\@@href}%
\providecommand \@@href[1]{\endgroup#1\@@endlink}%
\providecommand \@sanitize@url [0]{\catcode `\\12\catcode `\$12\catcode
  `\&12\catcode `\#12\catcode `\^12\catcode `\_12\catcode `\%12\relax}%
\providecommand \@@startlink[1]{}%
\providecommand \@@endlink[0]{}%
\providecommand \url  [0]{\begingroup\@sanitize@url \@url }%
\providecommand \@url [1]{\endgroup\@href {#1}{\urlprefix }}%
\providecommand \urlprefix  [0]{URL }%
\providecommand \Eprint [0]{\href }%
\providecommand \doibase [0]{http://dx.doi.org/}%
\providecommand \selectlanguage [0]{\@gobble}%
\providecommand \bibinfo  [0]{\@secondoftwo}%
\providecommand \bibfield  [0]{\@secondoftwo}%
\providecommand \translation [1]{[#1]}%
\providecommand \BibitemOpen [0]{}%
\providecommand \bibitemStop [0]{}%
\providecommand \bibitemNoStop [0]{.\EOS\space}%
\providecommand \EOS [0]{\spacefactor3000\relax}%
\providecommand \BibitemShut  [1]{\csname bibitem#1\endcsname}%
\let\auto@bib@innerbib\@empty
\bibitem [{\citenamefont {Aggarwal}\ \emph {et~al.}(2010)\citenamefont
  {Aggarwal} \emph{et~al.}}]{STAR2010}%
  \BibitemOpen
  \bibfield  {author} {\bibinfo {author} {\bibfnamefont {M.~M.}\ \bibnamefont
  {Aggarwal}} {\it{et~al.}} (\bibinfo {collaboration} {STAR Collaboration}),\ }\href@noop {}
 \ \Eprint {http://arxiv.org/abs/1007.2613}
  {arXiv: 1007.2613} \BibitemShut {NoStop}%
\bibitem [{sta()}]{starnote}%
  \BibitemOpen 
  {\bibinfo {title} {BES-II White Paper (STAR Note 2014)}},
  \href {https://drupal.star.bnl.gov/STAR/starnotes/public/sn0598.}{https://drupal.star .bnl.gov/STAR/starnotes/public/sn0598.} \BibitemShut
  {NoStop}%
\bibitem [{\citenamefont {Busza}\ \emph {et~al.}(2018)\citenamefont {Busza},
  \citenamefont {Rajagopal},\ and\ \citenamefont {van~der Schee}}]{Busza2018}%
  \BibitemOpen
  \bibfield  {author} {\bibinfo {author} {\bibfnamefont {W.}~\bibnamefont
  {Busza}}, \bibinfo {author} {\bibfnamefont {K.}~\bibnamefont {Rajagopal}}, \
  and\ \bibinfo {author} {\bibfnamefont {W.}~\bibnamefont {van~der Schee}},\
  }\href {\doibase 10.1146/annurev-nucl-101917-020852} {\bibfield  {journal}
  {\bibinfo  {journal} {Annu. Rev. Nucl. Part. Sci.}\
  }\textbf {\bibinfo {volume} {68}},\ \bibinfo {pages} {339} (\bibinfo {year}
  {2018})}\BibitemShut {NoStop}%
  \bibitem [{\citenamefont {Adam}\ \emph {et~al.}(2021)\citenamefont {Adam} \emph
  {et~al.}}]{STAR20202S}%
  \BibitemOpen
  \bibfield  {author} {\bibinfo {author} {\bibfnamefont {J.}~\bibnamefont
  {Adam}} {\it{et~al.}} (\bibinfo {collaboration} {STAR Collaboration}),\ }\href {\doibase
  10.1103/PhysRevLett.126.092301} {\bibfield  {journal} {\bibinfo  {journal}
  {Phys. Rev. Lett.}\ }\textbf {\bibinfo {volume} {126}},\ \bibinfo {pages}
  {092301} (\bibinfo {year} {2021})} \BibitemShut
  {NoStop}%
\bibitem [{\citenamefont {Abdallah}\ \emph {et~al.}(2021)\citenamefont
  {Abdallah} \emph {et~al.}}]{STAR2021}%
  \BibitemOpen
  \bibfield  {author} {\bibinfo {author} {\bibfnamefont {M.}~\bibnamefont
  {Abdallah}} {\it {et~al.}} (\bibinfo {collaboration} {STAR Collaboration}),\ }\href@noop {}
  \ \Eprint {http://arxiv.org/abs/2101.12413}
  {arXiv:2101.12413} \BibitemShut {NoStop}%
\bibitem [{\citenamefont {Letessier}\ and\ \citenamefont
  {Rafelski}(2002)}]{letessier_rafelski_2002}%
  \BibitemOpen
  \bibfield  {author} {\bibinfo {author} {\bibfnamefont {J.}~\bibnamefont
  {Letessier}}\ and\ \bibinfo {author} {\bibfnamefont {J.}~\bibnamefont
  {Rafelski}},\ }\href {\doibase 10.1017/CBO9780511534997} { {\bibinfo
  {title} {\it Hadrons and Quark–Gluon Plasma}}},\ Cambridge Monographs on
  Particle Physics, Nuclear Physics and Cosmology\ (\bibinfo  {publisher}
  {Cambridge University Press, Cambridge, England},\ \bibinfo {year} {2002})\BibitemShut {NoStop}%
\bibitem [{\citenamefont {Aoki}\ \emph {et~al.}(2006)\citenamefont {Aoki},
  \citenamefont {Endr{\H{o}}di}, \citenamefont {Fodor}, \citenamefont {Katz},\
  and\ \citenamefont {Szab{\'o}}}]{Aoki2006sf}%
  \BibitemOpen
  \bibfield  {author} {\bibinfo {author} {\bibfnamefont {Y.}~\bibnamefont
  {Aoki}}, \bibinfo {author} {\bibfnamefont {G.}~\bibnamefont {Endr{\H{o}}di}},
  \bibinfo {author} {\bibfnamefont {Z.}~\bibnamefont {Fodor}}, \bibinfo
  {author} {\bibfnamefont {S.~D.}\ \bibnamefont {Katz}}, \ and\ \bibinfo
  {author} {\bibfnamefont {K.~K.}\ \bibnamefont {Szab{\'o}}},\ }\href {\doibase
  10.1038/nature05120} {\bibfield  {journal} {\bibinfo  {journal} {Nature (London)}\
  }\textbf {\bibinfo {volume} {443}},\ \bibinfo {pages} {675} (\bibinfo {year}
  {2006})}\BibitemShut {NoStop}%
\bibitem [{\citenamefont {Bazavov}\ \emph {et~al.}(2012)\citenamefont {Bazavov}
  \emph {et~al.}}]{Bazavov2012el}%
  \BibitemOpen
  \bibfield  {author} {\bibinfo {author} {\bibfnamefont {A.}~\bibnamefont
  {Bazavov}} { \it{et~al.}} (\bibinfo {collaboration} {HotQCD Collaboration}),\ }\href
  {\doibase 10.1103/PhysRevD.85.054503} {\bibfield  {journal} {\bibinfo
  {journal} {Phys. Rev. D}\ }\textbf {\bibinfo {volume} {85}},\ \bibinfo
  {pages} {054503} (\bibinfo {year} {2012})}\BibitemShut {NoStop}%
\bibitem [{\citenamefont {Fukushima}\ and\ \citenamefont
  {Sasaki}(2013)}]{FUKUSHIMA201399}%
  \BibitemOpen
  \bibfield  {author} {\bibinfo {author} {\bibfnamefont {K.}~\bibnamefont
  {Fukushima}}\ and\ \bibinfo {author} {\bibfnamefont {C.}~\bibnamefont
  {Sasaki}},\ }\href {\doibase https://doi.org/10.1016/j.ppnp.2013.05.003}
  {\bibfield  {journal} {\bibinfo  {journal} {Prog. Part. Nucl.
  Phys.}\ }\textbf {\bibinfo {volume} {72}},\ \bibinfo {pages} {99} (\bibinfo
  {year} {2013})}\BibitemShut {NoStop}%
\bibitem [{\citenamefont {Bzdak}\ \emph {et~al.}(2020)\citenamefont {Bzdak},
  \citenamefont {Esumi}, \citenamefont {Koch}, \citenamefont {Liao},
  \citenamefont {Stephanov},\ and\ \citenamefont {Xu}}]{BZDAK20201}%
  \BibitemOpen
  \bibfield  {author} {\bibinfo {author} {\bibfnamefont {A.}~\bibnamefont
  {Bzdak}}, \bibinfo {author} {\bibfnamefont {S.}~\bibnamefont {Esumi}},
  \bibinfo {author} {\bibfnamefont {V.}~\bibnamefont {Koch}}, \bibinfo {author}
  {\bibfnamefont {J.}~\bibnamefont {Liao}}, \bibinfo {author} {\bibfnamefont
  {M.}~\bibnamefont {Stephanov}}, \ and\ \bibinfo {author} {\bibfnamefont
  {N.}~\bibnamefont {Xu}},\ }\href {\doibase
  https://doi.org/10.1016/j.physrep.2020.01.005} {\bibfield  {journal}
  {\bibinfo  {journal} {Phys. Rep.}\ }\textbf {\bibinfo {volume} {853}},\
  \bibinfo {pages} {1} (\bibinfo {year} {2020})}\BibitemShut {NoStop}%
\bibitem [{\citenamefont {Adamczyk}\ \emph {et~al.}(2017)\citenamefont
  {Adamczyk} \emph {et~al.}}]{Adamczyk2017}%
  \BibitemOpen
  \bibfield  {author} {\bibinfo {author} {\bibfnamefont {L.}~\bibnamefont
  {Adamczyk}}  {\it et~al.} (\bibinfo {collaboration} {STAR Collaboration}),\ }\href
  {\doibase 10.1103/PhysRevC.96.044904} {\bibfield  {journal} {\bibinfo
  {journal} {Phys. Rev. C}\ }\textbf {\bibinfo {volume} {96}},\ \bibinfo
  {pages} {044904} (\bibinfo {year} {2017})}\BibitemShut {NoStop}%
\bibitem [{\citenamefont {Abelev}\ \emph {et~al.}(2013)\citenamefont {Abelev}
  \emph {et~al.}}]{Alice2013prc}%
  \BibitemOpen
  \bibfield  {author} {\bibinfo {author} {\bibfnamefont {B.}~\bibnamefont
  {Abelev}}  {\it et~al.} (\bibinfo {collaboration} {ALICE Collaboration}),\
  }\href {\doibase 10.1103/PhysRevC.88.044910} {\bibfield  {journal} {\bibinfo
  {journal} {Phys. Rev. C}\ }\textbf {\bibinfo {volume} {88}},\ \bibinfo
  {pages} {044910} (\bibinfo {year} {2013})}\BibitemShut {NoStop}%
\bibitem [{\citenamefont {Abelev}\ \emph {et~al.}(2012)\citenamefont {Abelev}
  \emph {et~al.}}]{Alice2013prl}%
  \BibitemOpen
  \bibfield  {author} {\bibinfo {author} {\bibfnamefont {B.}~\bibnamefont
  {Abelev}}  {\it et~al.} (\bibinfo {collaboration} {ALICE Collaboration}),\
  }\href {\doibase 10.1103/PhysRevLett.109.252301} {\bibfield  {journal}
  {\bibinfo  {journal} {Phys. Rev. Lett.}\ }\textbf {\bibinfo {volume} {109}},\
  \bibinfo {pages} {252301} (\bibinfo {year} {2012})}\BibitemShut {NoStop}%
\bibitem [{\citenamefont {Andronic}\ \emph {et~al.}(2018)\citenamefont
  {Andronic}, \citenamefont {Braun-Munzinger}, \citenamefont {Redlich},\ and\
  \citenamefont {Stachel}}]{Andronic2018}%
  \BibitemOpen
  \bibfield  {author} {\bibinfo {author} {\bibfnamefont {A.}~\bibnamefont
  {Andronic}}, \bibinfo {author} {\bibfnamefont {P.}~\bibnamefont
  {Braun-Munzinger}}, \bibinfo {author} {\bibfnamefont {K.}~\bibnamefont
  {Redlich}}, \ and\ \bibinfo {author} {\bibfnamefont {J.}~\bibnamefont
  {Stachel}},\ }\href {\doibase 10.1038/s41586-018-0491-6} {\bibfield
  {journal} {\bibinfo  {journal} {Nature (London)}\ }\textbf {\bibinfo {volume} {561}},\
  \bibinfo {pages} {321} (\bibinfo {year} {2018})}\BibitemShut {NoStop}%
\bibitem [{\citenamefont {Luo}\ \emph {et~al.}(2020)\citenamefont {Luo},
  \citenamefont {Shi}, \citenamefont {Xu},\ and\ \citenamefont
  {Zhang}}]{Luo2020}%
  \BibitemOpen
  \bibfield  {author} {\bibinfo {author} {\bibfnamefont {X.}~\bibnamefont
  {Luo}}, \bibinfo {author} {\bibfnamefont {S.}~\bibnamefont {Shi}}, \bibinfo
  {author} {\bibfnamefont {N.}~\bibnamefont {Xu}}, \ and\ \bibinfo {author}
  {\bibfnamefont {Y.}~\bibnamefont {Zhang}},\ }\href {\doibase
  10.3390/particles3020022} {\bibfield  {journal} {\bibinfo  {journal}
  {Particles}\ }\textbf {\bibinfo {volume} {3}},\ \bibinfo {pages} {278}
  (\bibinfo {year} {2020})}\BibitemShut {NoStop}%
\bibitem [{\citenamefont {Fukushima}\ \emph {et~al.}(2020)\citenamefont
  {Fukushima}, \citenamefont {Mohanty},\ and\ \citenamefont
  {Xu}}]{Fukushima:2020yzx}%
  \BibitemOpen
  \bibfield  {author} {\bibinfo {author} {\bibfnamefont {K.}~\bibnamefont
  {Fukushima}}, \bibinfo {author} {\bibfnamefont {B.}~\bibnamefont {Mohanty}},
  \ and\ \bibinfo {author} {\bibfnamefont {N.}~\bibnamefont {Xu}},\ }\href {\doibase
  10.1007/s43673-021-00002-7} {\bibfield  {journal} {\bibinfo  {journal}
  {AAPPS Bull.}\ }\textbf {\bibinfo {volume} {31}},\ \bibinfo {pages} {1}
  (\bibinfo {year} {2021})}\BibitemShut {NoStop}%
\bibitem [{\citenamefont {Bazavov}\ \emph {et~al.}(2014)\citenamefont {Bazavov}
  \emph {et~al.}}]{Bazavov2014}%
  \BibitemOpen
  \bibfield  {author} {\bibinfo {author} {\bibfnamefont {A.}~\bibnamefont
  {Bazavov}}  {\it et~al.} (\bibinfo {collaboration} {HotQCD Collaboration}),\ }\href
  {\doibase 10.1103/PhysRevD.90.094503} {\bibfield  {journal} {\bibinfo
  {journal} {Phys. Rev. D}\ }\textbf {\bibinfo {volume} {90}},\ \bibinfo
  {pages} {094503} (\bibinfo {year} {2014})}\BibitemShut {NoStop}%
\bibitem [{\citenamefont {Kaczmarek}\ \emph {et~al.}(2011)\citenamefont
  {Kaczmarek}, \citenamefont {Karsch}, \citenamefont {Laermann}, \citenamefont
  {Miao}, \citenamefont {Mukherjee}, \citenamefont {Petreczky}, \citenamefont
  {Schmidt}, \citenamefont {Soeldner},\ and\ \citenamefont
  {Unger}}]{Kaczmarek:2011zz}%
  \BibitemOpen
  \bibfield  {author} {\bibinfo {author} {\bibfnamefont {O.}~\bibnamefont
  {Kaczmarek}}, \bibinfo {author} {\bibfnamefont {F.}~\bibnamefont {Karsch}},
  \bibinfo {author} {\bibfnamefont {E.}~\bibnamefont {Laermann}}, \bibinfo
  {author} {\bibfnamefont {C.}~\bibnamefont {Miao}}, \bibinfo {author}
  {\bibfnamefont {S.}~\bibnamefont {Mukherjee}}, \bibinfo {author}
  {\bibfnamefont {P.}~\bibnamefont {Petreczky}}, \bibinfo {author}
  {\bibfnamefont {C.}~\bibnamefont {Schmidt}}, \bibinfo {author} {\bibfnamefont
  {W.}~\bibnamefont {Soeldner}}, \ and\ \bibinfo {author} {\bibfnamefont
  {W.}~\bibnamefont {Unger}},\ }\href {\doibase 10.1103/PhysRevD.83.014504}
  {\bibfield  {journal} {\bibinfo  {journal} {Phys. Rev. D}\ }\textbf {\bibinfo
  {volume} {83}},\ \bibinfo {pages} {014504} (\bibinfo {year}
  {2011})}\BibitemShut {NoStop}%
\bibitem [{\citenamefont {Bazavov}\ \emph {et~al.}(2019)\citenamefont {Bazavov}
  \emph {et~al.}}]{Bazavov:2018mes}%
  \BibitemOpen
  \bibfield  {author} {\bibinfo {author} {\bibfnamefont {A.}~\bibnamefont
  {Bazavov}}  {\it et~al.} (\bibinfo {collaboration} {HotQCD Collaboration}),\ }\href
  {\doibase 10.1016/j.physletb.2019.05.013} {\bibfield  {journal} {\bibinfo
  {journal} {Phys. Lett. B}\ }\textbf {\bibinfo {volume} {795}},\ \bibinfo
  {pages} {15} (\bibinfo {year} {2019})} \BibitemShut
  {NoStop}%
\bibitem [{\citenamefont {Fu}\ \emph {et~al.}(2021)\citenamefont {Fu},
  \citenamefont {Luo}, \citenamefont {Pawlowski}, \citenamefont {Rennecke},
  \citenamefont {Wen},\ and\ \citenamefont {Yin}}]{Fu:2021oaw}%
  \BibitemOpen
  \bibfield  {author} {\bibinfo {author} {\bibfnamefont {W.-j.}\ \bibnamefont
  {Fu}}, \bibinfo {author} {\bibfnamefont {X.}~\bibnamefont {Luo}}, \bibinfo
  {author} {\bibfnamefont {J.~M.}\ \bibnamefont {Pawlowski}}, \bibinfo {author}
  {\bibfnamefont {F.}~\bibnamefont {Rennecke}}, \bibinfo {author}
  {\bibfnamefont {R.}~\bibnamefont {Wen}}, \ and\ \bibinfo {author}
  {\bibfnamefont {S.}~\bibnamefont {Yin}},\ }\href@noop {} \ \Eprint {http://arxiv.org/abs/2101.06035} {arXiv:2101.06035} \BibitemShut {NoStop}%
\bibitem [{\citenamefont {Berdnikov}\ and\ \citenamefont
  {Rajagopal}(2000)}]{Berdnikov2000}%
  \BibitemOpen
  \bibfield  {author} {\bibinfo {author} {\bibfnamefont {B.}~\bibnamefont
  {Berdnikov}}\ and\ \bibinfo {author} {\bibfnamefont {K.}~\bibnamefont
  {Rajagopal}},\ }\href {\doibase 10.1103/PhysRevD.61.105017} {\bibfield
  {journal} {\bibinfo  {journal} {Phys. Rev. D}\ }\textbf {\bibinfo {volume}
  {61}},\ \bibinfo {pages} {105017} (\bibinfo {year} {2000})}\BibitemShut
  {NoStop}%
\bibitem [{\citenamefont {Mukherjee}\ \emph {et~al.}(2015)\citenamefont
  {Mukherjee}, \citenamefont {Venugopalan},\ and\ \citenamefont
  {Yin}}]{Mukherjee2015}%
  \BibitemOpen
  \bibfield  {author} {\bibinfo {author} {\bibfnamefont {S.}~\bibnamefont
  {Mukherjee}}, \bibinfo {author} {\bibfnamefont {R.}~\bibnamefont
  {Venugopalan}}, \ and\ \bibinfo {author} {\bibfnamefont {Y.}~\bibnamefont
  {Yin}},\ }\href {\doibase 10.1103/PhysRevC.92.034912} {\bibfield  {journal}
  {\bibinfo  {journal} {Phys. Rev. C}\ }\textbf {\bibinfo {volume} {92}},\
  \bibinfo {pages} {034912} (\bibinfo {year} {2015})}\BibitemShut {NoStop}%
\bibitem [{\citenamefont {Mukherjee}\ \emph {et~al.}(2016)\citenamefont
  {Mukherjee}, \citenamefont {Venugopalan},\ and\ \citenamefont
  {Yin}}]{Swagato2016}%
  \BibitemOpen
  \bibfield  {author} {\bibinfo {author} {\bibfnamefont {S.}~\bibnamefont
  {Mukherjee}}, \bibinfo {author} {\bibfnamefont {R.}~\bibnamefont
  {Venugopalan}}, \ and\ \bibinfo {author} {\bibfnamefont {Y.}~\bibnamefont
  {Yin}},\ }\href {\doibase 10.1103/PhysRevLett.117.222301} {\bibfield
  {journal} {\bibinfo  {journal} {Phys. Rev. Lett.}\ }\textbf {\bibinfo
  {volume} {117}},\ \bibinfo {pages} {222301} (\bibinfo {year}
  {2016})}\BibitemShut {NoStop}%
\bibitem [{\citenamefont {Jiang}\ \emph
  {et~al.}(2017{\natexlab{a}})\citenamefont {Jiang}, \citenamefont {Wu},\ and\
  \citenamefont {Song}}]{Jiangvr2017}%
  \BibitemOpen
  \bibfield  {author} {\bibinfo {author} {\bibfnamefont {L.}~\bibnamefont
  {Jiang}}, \bibinfo {author} {\bibfnamefont {S.}~\bibnamefont {Wu}}, \ and\
  \bibinfo {author} {\bibfnamefont {H.}~\bibnamefont {Song}},\ }\href {\doibase
  10.1016/j.nuclphysa.2017.06.047} {\bibfield  {journal} {\bibinfo  {journal}
  {Nucl. Phys.}\ }\textbf {\bibinfo {volume} {A967}},\ \bibinfo {pages} {441}
  (\bibinfo {year} {2017}{\natexlab{a}})}\BibitemShut {NoStop}%
\bibitem [{\citenamefont {Wu}\ \emph {et~al.}(2019)\citenamefont {Wu},
  \citenamefont {Wu},\ and\ \citenamefont {Song}}]{SJWu2019}%
  \BibitemOpen
  \bibfield  {author} {\bibinfo {author} {\bibfnamefont {S.}~\bibnamefont
  {Wu}}, \bibinfo {author} {\bibfnamefont {Z.}~\bibnamefont {Wu}}, \ and\
  \bibinfo {author} {\bibfnamefont {H.}~\bibnamefont {Song}},\ }\href {\doibase
  10.1103/PhysRevC.99.064902} {\bibfield  {journal} {\bibinfo  {journal} {Phys.
  Rev. C}\ }\textbf {\bibinfo {volume} {99}},\ \bibinfo {pages} {064902}
  (\bibinfo {year} {2019})}\BibitemShut {NoStop}%
\bibitem [{\citenamefont {Stephanov}(2009)}]{Stephanov2009}%
  \BibitemOpen
  \bibfield  {author} {\bibinfo {author} {\bibfnamefont {M.~A.}\ \bibnamefont
  {Stephanov}},\ }\href {\doibase 10.1103/PhysRevLett.102.032301} {\bibfield
  {journal} {\bibinfo  {journal} {Phys. Rev. Lett.}\ }\textbf {\bibinfo
  {volume} {102}},\ \bibinfo {pages} {032301} (\bibinfo {year}
  {2009})}\BibitemShut {NoStop}%
\bibitem [{\citenamefont {{Jeon}}\ and\ \citenamefont
  {{Koch}}(2004)}]{Jeon2004}%
  \BibitemOpen
  \bibfield  {author} {\bibinfo {author} {\bibfnamefont {S.}~\bibnamefont
  {{Jeon}}}\ and\ \bibinfo {author} {\bibfnamefont {V.}~\bibnamefont
  {{Koch}}},\ }in\ \href {\doibase 10.1142/5029}{ {\bibinfo {booktitle} {\it Quark-Gluon
  Plasma 3}}},\ \bibinfo {editor} {edited by\ \bibinfo {editor} {\bibfnamefont
  {R.~C.}\ \bibnamefont {{Hwa}}}\ and\ \bibinfo {editor} {\bibfnamefont
  {X.~N.}\ \bibnamefont {{Wang}}}}\ (\bibinfo {publisher}
  {World Scientific, Singapore,} \bibinfo {year} {2004})\ p.\ \bibinfo
  {pages} {430} 
  \BibitemShut {NoStop}%
\bibitem [{\citenamefont {Stephanov}\ \emph {et~al.}(1999)\citenamefont
  {Stephanov}, \citenamefont {Rajagopal},\ and\ \citenamefont
  {Shuryak}}]{Stephanov1999}%
  \BibitemOpen
  \bibfield  {author} {\bibinfo {author} {\bibfnamefont {M.}~\bibnamefont
  {Stephanov}}, \bibinfo {author} {\bibfnamefont {K.}~\bibnamefont
  {Rajagopal}}, \ and\ \bibinfo {author} {\bibfnamefont {E.}~\bibnamefont
  {Shuryak}},\ }\href {\doibase 10.1103/PhysRevD.60.114028} {\bibfield
  {journal} {\bibinfo  {journal} {Phys. Rev. D}\ }\textbf {\bibinfo {volume}
  {60}},\ \bibinfo {pages} {114028} (\bibinfo {year} {1999})}\BibitemShut
  {NoStop}%
\bibitem [{\citenamefont {Huovinen}\ \emph {et~al.}(2001)\citenamefont
  {Huovinen}, \citenamefont {Kolb}, \citenamefont {Heinz}, \citenamefont
  {Ruuskanen},\ and\ \citenamefont {Voloshin}}]{Heinz2001}%
  \BibitemOpen
  \bibfield  {author} {\bibinfo {author} {\bibfnamefont {P.}~\bibnamefont
  {Huovinen}}, \bibinfo {author} {\bibfnamefont {P.}~\bibnamefont {Kolb}},
  \bibinfo {author} {\bibfnamefont {U.}~\bibnamefont {Heinz}}, \bibinfo
  {author} {\bibfnamefont {P.}~\bibnamefont {Ruuskanen}}, \ and\ \bibinfo
  {author} {\bibfnamefont {S.}~\bibnamefont {Voloshin}},\ }\href {\doibase
  https://doi.org/10.1016/S0370-2693(01)00219-2} {\bibfield  {journal}
  {\bibinfo  {journal} {Phys. Lett. B}\ }\textbf {\bibinfo {volume}
  {503}},\ \bibinfo {pages} {58} (\bibinfo {year} {2001})}\BibitemShut
  {NoStop}%
\bibitem [{\citenamefont {Luzum}\ and\ \citenamefont
  {Romatschke}(2008)}]{Luzum2008}%
  \BibitemOpen
  \bibfield  {author} {\bibinfo {author} {\bibfnamefont {M.}~\bibnamefont
  {Luzum}}\ and\ \bibinfo {author} {\bibfnamefont {P.}~\bibnamefont
  {Romatschke}},\ }\href {\doibase 10.1103/PhysRevC.78.034915} {\bibfield
  {journal} {\bibinfo  {journal} {Phys. Rev. C}\ }\textbf {\bibinfo {volume}
  {78}},\ \bibinfo {pages} {034915} (\bibinfo {year} {2008})}\BibitemShut
  {NoStop}%
\bibitem [{\citenamefont {Gale}\ \emph {et~al.}(2013)\citenamefont {Gale},
  \citenamefont {Jeon}, \citenamefont {Schenke}, \citenamefont {Tribedy},\ and\
  \citenamefont {Venugopalan}}]{Gale2013}%
  \BibitemOpen
  \bibfield  {author} {\bibinfo {author} {\bibfnamefont {C.}~\bibnamefont
  {Gale}}, \bibinfo {author} {\bibfnamefont {S.}~\bibnamefont {Jeon}}, \bibinfo
  {author} {\bibfnamefont {B.}~\bibnamefont {Schenke}}, \bibinfo {author}
  {\bibfnamefont {P.}~\bibnamefont {Tribedy}}, \ and\ \bibinfo {author}
  {\bibfnamefont {R.}~\bibnamefont {Venugopalan}},\ }\href {\doibase
  10.1103/PhysRevLett.110.012302} {\bibfield  {journal} {\bibinfo  {journal}
  {Phys. Rev. Lett.}\ }\textbf {\bibinfo {volume} {110}},\ \bibinfo {pages}
  {012302} (\bibinfo {year} {2013})}\BibitemShut {NoStop}%
\bibitem [{\citenamefont {Song}\ and\ \citenamefont {Heinz}(2008)}]{Song2007}%
  \BibitemOpen
  \bibfield  {author} {\bibinfo {author} {\bibfnamefont {H.}~\bibnamefont
  {Song}}\ and\ \bibinfo {author} {\bibfnamefont {U.}~\bibnamefont {Heinz}},\
  }\href {\doibase 10.1103/PhysRevC.77.064901} {\bibfield  {journal} {\bibinfo
  {journal} {Phys. Rev. C}\ }\textbf {\bibinfo {volume} {77}},\ \bibinfo
  {pages} {064901} (\bibinfo {year} {2008})}\BibitemShut {NoStop}%
\bibitem [{\citenamefont {Du}\ and\ \citenamefont {Heinz}(2020)}]{DU2020hy}%
  \BibitemOpen
  \bibfield  {author} {\bibinfo {author} {\bibfnamefont {L.}~\bibnamefont
  {Du}}\ and\ \bibinfo {author} {\bibfnamefont {U.}~\bibnamefont {Heinz}},\
  }\href {\doibase https://doi.org/10.1016/j.cpc.2019.107090} {\bibfield
  {journal} {\bibinfo  {journal} {Comput. Phys. Commun.}\ }\textbf
  {\bibinfo {volume} {251}},\ \bibinfo {pages} {107090} (\bibinfo {year}
  {2020})}\BibitemShut {NoStop}%
\bibitem [{\citenamefont {Shen}\ and\ \citenamefont
  {Alzhrani}(2020)}]{Shen2020fi}%
  \BibitemOpen
  \bibfield  {author} {\bibinfo {author} {\bibfnamefont {C.}~\bibnamefont
  {Shen}}\ and\ \bibinfo {author} {\bibfnamefont {S.}~\bibnamefont
  {Alzhrani}},\ }\href {\doibase 10.1103/PhysRevC.102.014909} {\bibfield
  {journal} {\bibinfo  {journal} {Phys. Rev. C}\ }\textbf {\bibinfo {volume}
  {102}},\ \bibinfo {pages} {014909} (\bibinfo {year} {2020})}\BibitemShut
  {NoStop}%
\bibitem [{\citenamefont {Skokov}\ \emph {et~al.}(2010)\citenamefont {Skokov},
  \citenamefont {Friman}, \citenamefont {Nakano}, \citenamefont {Redlich},\
  and\ \citenamefont {Schaefer}}]{Skokov2010sj}%
  \BibitemOpen
  \bibfield  {author} {\bibinfo {author} {\bibfnamefont {V.}~\bibnamefont
  {Skokov}}, \bibinfo {author} {\bibfnamefont {B.}~\bibnamefont {Friman}},
  \bibinfo {author} {\bibfnamefont {E.}~\bibnamefont {Nakano}}, \bibinfo
  {author} {\bibfnamefont {K.}~\bibnamefont {Redlich}}, \ and\ \bibinfo
  {author} {\bibfnamefont {B.-J.}\ \bibnamefont {Schaefer}},\ }\href {\doibase
  10.1103/PhysRevD.82.034029} {\bibfield  {journal} {\bibinfo  {journal} {Phys.
  Rev. D}\ }\textbf {\bibinfo {volume} {82}},\ \bibinfo {pages} {034029}
  (\bibinfo {year} {2010})}\BibitemShut {NoStop}%
\bibitem [{\citenamefont {Roberts}\ and\ \citenamefont
  {Williams}(1994)}]{ROBERTS1994477}%
  \BibitemOpen
  \bibfield  {author} {\bibinfo {author} {\bibfnamefont {C.~D.}\ \bibnamefont
  {Roberts}}\ and\ \bibinfo {author} {\bibfnamefont {A.~G.}\ \bibnamefont
  {Williams}},\ }\href {\doibase https://doi.org/10.1016/0146-6410(94)90049-3}
  {\bibfield  {journal} {\bibinfo  {journal} {Prog. Part. Nucl.
  Phys.}\ }\textbf {\bibinfo {volume} {33}},\ \bibinfo {pages} {477}
  (\bibinfo {year} {1994})}\BibitemShut {NoStop}%
\bibitem [{\citenamefont {Jungnickel}\ and\ \citenamefont
  {Wetterich}(1996)}]{Jungnickel:1995fp}%
  \BibitemOpen
  \bibfield  {author} {\bibinfo {author} {\bibfnamefont {D.~U.}\ \bibnamefont
  {Jungnickel}}\ and\ \bibinfo {author} {\bibfnamefont {C.}~\bibnamefont
  {Wetterich}},\ }\href {\doibase 10.1103/PhysRevD.53.5142} {\bibfield
  {journal} {\bibinfo  {journal} {Phys. Rev. D}\ }\textbf {\bibinfo {volume}
  {53}},\ \bibinfo {pages} {5142} (\bibinfo {year} {1996})}\BibitemShut
  {NoStop}%
\bibitem [{\citenamefont {Schaefer}\ and\ \citenamefont
  {Wambach}(2008)}]{Schaefer:2006sr}%
  \BibitemOpen
  \bibfield  {author} {\bibinfo {author} {\bibfnamefont {B.-J.}\ \bibnamefont
  {Schaefer}}\ and\ \bibinfo {author} {\bibfnamefont {J.}~\bibnamefont
  {Wambach}},\ }\href {\doibase 10.1134/S1063779608070083} {\bibfield
  {journal} {\bibinfo  {journal} {Phys. Part. Nucl.}\ }\textbf {\bibinfo
  {volume} {39}},\ \bibinfo {pages} {1025} (\bibinfo {year}
  {2008})}\BibitemShut {NoStop}%
\bibitem [{\citenamefont {Qin}\ \emph {et~al.}(2011)\citenamefont {Qin},
  \citenamefont {Chang}, \citenamefont {Chen}, \citenamefont {Liu},\ and\
  \citenamefont {Roberts}}]{qin2011}%
  \BibitemOpen
  \bibfield  {author} {\bibinfo {author} {\bibfnamefont {S.-X.}\ \bibnamefont
  {Qin}}, \bibinfo {author} {\bibfnamefont {L.}~\bibnamefont {Chang}}, \bibinfo
  {author} {\bibfnamefont {H.}~\bibnamefont {Chen}}, \bibinfo {author}
  {\bibfnamefont {Y.-X.}\ \bibnamefont {Liu}}, \ and\ \bibinfo {author}
  {\bibfnamefont {C.~D.}\ \bibnamefont {Roberts}},\ }\href {\doibase
  10.1103/PhysRevLett.106.172301} {\bibfield  {journal} {\bibinfo  {journal}
  {Phys. Rev. Lett.}\ }\textbf {\bibinfo {volume} {106}},\ \bibinfo {pages}
  {172301} (\bibinfo {year} {2011})}\BibitemShut {NoStop}%
\bibitem [{\citenamefont {Jiang}\ \emph {et~al.}(2013)\citenamefont {Jiang},
  \citenamefont {Xin}, \citenamefont {Wang}, \citenamefont {Qin},\ and\
  \citenamefont {Liu}}]{Jiang2013}%
  \BibitemOpen
  \bibfield  {author} {\bibinfo {author} {\bibfnamefont {L.-J.}\ \bibnamefont
  {Jiang}}, \bibinfo {author} {\bibfnamefont {X.-Y.}\ \bibnamefont {Xin}},
  \bibinfo {author} {\bibfnamefont {K.-L.}\ \bibnamefont {Wang}}, \bibinfo
  {author} {\bibfnamefont {S.-X.}\ \bibnamefont {Qin}}, \ and\ \bibinfo
  {author} {\bibfnamefont {Y.-X.}\ \bibnamefont {Liu}},\ }\href {\doibase
  10.1103/PhysRevD.88.016008} {\bibfield  {journal} {\bibinfo  {journal} {Phys.
  Rev. D}\ }\textbf {\bibinfo {volume} {88}},\ \bibinfo {pages} {016008}
  (\bibinfo {year} {2013})}\BibitemShut {NoStop}%
\bibitem [{\citenamefont {Fukushima}\ and\ \citenamefont
  {Hatsuda}(2010)}]{Fukushima_2010}%
  \BibitemOpen
  \bibfield  {author} {\bibinfo {author} {\bibfnamefont {K.}~\bibnamefont
  {Fukushima}}\ and\ \bibinfo {author} {\bibfnamefont {T.}~\bibnamefont
  {Hatsuda}},\ }\href {\doibase 10.1088/0034-4885/74/1/014001} {\bibfield
  {journal} {\bibinfo  {journal} {Rep.  Prog. Phys.}\ }\textbf
  {\bibinfo {volume} {74}},\ \bibinfo {pages} {014001} (\bibinfo {year}
  {2010})}\BibitemShut {NoStop}%
\bibitem [{\citenamefont {Schaefer}\ and\ \citenamefont
  {Wambach}(2005)}]{SCHAEFER2005479}%
  \BibitemOpen
  \bibfield  {author} {\bibinfo {author} {\bibfnamefont {B.-J.}\ \bibnamefont
  {Schaefer}}\ and\ \bibinfo {author} {\bibfnamefont {J.}~\bibnamefont
  {Wambach}},\ }\href {\doibase
  https://doi.org/10.1016/j.nuclphysa.2005.04.012} {\bibfield  {journal}
  {\bibinfo  {journal} {Nucl. Phys.}\ }\textbf {\bibinfo {volume}
  {A757}},\ \bibinfo {pages} {479} (\bibinfo {year} {2005})}\BibitemShut
  {NoStop}%
\bibitem [{\citenamefont {Scavenius}\ \emph {et~al.}(2001)\citenamefont
  {Scavenius}, \citenamefont {M\'ocsy}, \citenamefont {Mishustin},\ and\
  \citenamefont {Rischke}}]{Scavenius2001}%
  \BibitemOpen
  \bibfield  {author} {\bibinfo {author} {\bibfnamefont {O.}~\bibnamefont
  {Scavenius}}, \bibinfo {author} {\bibfnamefont {A.}~\bibnamefont {M\'ocsy}},
  \bibinfo {author} {\bibfnamefont {I.~N.}\ \bibnamefont {Mishustin}}, \ and\
  \bibinfo {author} {\bibfnamefont {D.~H.}\ \bibnamefont {Rischke}},\ }\href
  {\doibase 10.1103/PhysRevC.64.045202} {\bibfield  {journal} {\bibinfo
  {journal} {Phys. Rev. C}\ }\textbf {\bibinfo {volume} {64}},\ \bibinfo
  {pages} {045202} (\bibinfo {year} {2001})}\BibitemShut {NoStop}%
\bibitem [{\citenamefont {Paech}\ \emph {et~al.}(2003)\citenamefont {Paech},
  \citenamefont {St\"ocker},\ and\ \citenamefont {Dumitru}}]{Paech2003}%
  \BibitemOpen
  \bibfield  {author} {\bibinfo {author} {\bibfnamefont {K.}~\bibnamefont
  {Paech}}, \bibinfo {author} {\bibfnamefont {H.}~\bibnamefont {St\"ocker}}, \
  and\ \bibinfo {author} {\bibfnamefont {A.}~\bibnamefont {Dumitru}},\ }\href
  {\doibase 10.1103/PhysRevC.68.044907} {\bibfield  {journal} {\bibinfo
  {journal} {Phys. Rev. C}\ }\textbf {\bibinfo {volume} {68}},\ \bibinfo
  {pages} {044907} (\bibinfo {year} {2003})}\BibitemShut {NoStop}%
\bibitem [{\citenamefont {Jiang}\ \emph
  {et~al.}(2017{\natexlab{b}})\citenamefont {Jiang}, \citenamefont {Zheng},\
  and\ \citenamefont {Stoecker}}]{Jiang2017}%
  \BibitemOpen
  \bibfield  {author} {\bibinfo {author} {\bibfnamefont {L.}~\bibnamefont
  {Jiang}}, \bibinfo {author} {\bibfnamefont {J.-H.}\ \bibnamefont {Zheng}}, \
  and\ \bibinfo {author} {\bibfnamefont {H.}~\bibnamefont {Stoecker}},\
  }\href@noop {},\ \Eprint
  {http://arxiv.org/abs/1711.05339} {arXiv:1711.05339} \BibitemShut
  {NoStop}%
\bibitem [{\citenamefont {Pradeep}\ and\ \citenamefont
  {Stephanov}(2019)}]{Stephanov2019}%
  \BibitemOpen
  \bibfield  {author} {\bibinfo {author} {\bibfnamefont {M.~S.}\ \bibnamefont
  {Pradeep}}\ and\ \bibinfo {author} {\bibfnamefont {M.}~\bibnamefont
  {Stephanov}},\ }\href {\doibase 10.1103/PhysRevD.100.056003} {\bibfield
  {journal} {\bibinfo  {journal} {Phys. Rev. D}\ }\textbf {\bibinfo {volume}
  {100}},\ \bibinfo {pages} {056003} (\bibinfo {year} {2019})}\BibitemShut
  {NoStop}%
\bibitem [{\citenamefont {Nonaka}\ and\ \citenamefont
  {Asakawa}(2005)}]{Nonaka2005}%
  \BibitemOpen
  \bibfield  {author} {\bibinfo {author} {\bibfnamefont {C.}~\bibnamefont
  {Nonaka}}\ and\ \bibinfo {author} {\bibfnamefont {M.}~\bibnamefont
  {Asakawa}},\ }\href {\doibase 10.1103/PhysRevC.71.044904} {\bibfield
  {journal} {\bibinfo  {journal} {Phys. Rev. C}\ }\textbf {\bibinfo {volume}
  {71}},\ \bibinfo {pages} {044904} (\bibinfo {year} {2005})}\BibitemShut
  {NoStop}%
\bibitem [{\citenamefont {Stephanov}(2011)}]{Stephanov2011}%
  \BibitemOpen
  \bibfield  {author} {\bibinfo {author} {\bibfnamefont {M.~A.}\ \bibnamefont
  {Stephanov}},\ }\href {\doibase 10.1103/PhysRevLett.107.052301} {\bibfield
  {journal} {\bibinfo  {journal} {Phys. Rev. Lett.}\ }\textbf {\bibinfo
  {volume} {107}},\ \bibinfo {pages} {052301} (\bibinfo {year}
  {2011})}\BibitemShut {NoStop}%
\bibitem [{\citenamefont {Jiang}\ \emph
  {et~al.}(2016{\natexlab{a}})\citenamefont {Jiang}, \citenamefont {Li},\ and\
  \citenamefont {Song}}]{Jiang2016}%
  \BibitemOpen
  \bibfield  {author} {\bibinfo {author} {\bibfnamefont {L.}~\bibnamefont
  {Jiang}}, \bibinfo {author} {\bibfnamefont {P.}~\bibnamefont {Li}}, \ and\
  \bibinfo {author} {\bibfnamefont {H.}~\bibnamefont {Song}},\ }\href {\doibase
  10.1103/PhysRevC.94.024918} {\bibfield  {journal} {\bibinfo  {journal} {Phys.
  Rev. C}\ }\textbf {\bibinfo {volume} {94}},\ \bibinfo {pages} {024918}
  (\bibinfo {year} {2016}{\natexlab{a}})}\BibitemShut {NoStop}%
\bibitem [{\citenamefont {Jiang}\ \emph
  {et~al.}(2016{\natexlab{b}})\citenamefont {Jiang}, \citenamefont {Li},\ and\
  \citenamefont {Song}}]{JIANG2016360}%
  \BibitemOpen
  \bibfield  {author} {\bibinfo {author} {\bibfnamefont {L.}~\bibnamefont
  {Jiang}}, \bibinfo {author} {\bibfnamefont {P.}~\bibnamefont {Li}}, \ and\
  \bibinfo {author} {\bibfnamefont {H.}~\bibnamefont {Song}},\ }\href {\doibase
  https://doi.org/10.1016/j.nuclphysa.2016.01.034} {\bibfield  {journal}
  {\bibinfo  {journal} {Nucl. Phys.}\ }\textbf {\bibinfo {volume}
  {A956}},\ \bibinfo {pages} {360} (\bibinfo {year} {2016}{\natexlab{b}})},\
  \bibinfo {note} {the XXV International Conference on Ultrarelativistic
  Nucleus-Nucleus Collisions: Quark Matter 2015}\BibitemShut {NoStop}%
\bibitem [{\citenamefont {Herold}\ \emph {et~al.}(2016)\citenamefont {Herold},
  \citenamefont {Nahrgang}, \citenamefont {Yan},\ and\ \citenamefont
  {Kobdaj}}]{Herold2016prc}%
  \BibitemOpen
  \bibfield  {author} {\bibinfo {author} {\bibfnamefont {C.}~\bibnamefont
  {Herold}}, \bibinfo {author} {\bibfnamefont {M.}~\bibnamefont {Nahrgang}},
  \bibinfo {author} {\bibfnamefont {Y.}~\bibnamefont {Yan}}, \ and\ \bibinfo
  {author} {\bibfnamefont {C.}~\bibnamefont {Kobdaj}},\ }\href {\doibase
  10.1103/PhysRevC.93.021902} {\bibfield  {journal} {\bibinfo  {journal} {Phys.
  Rev. C}\ }\textbf {\bibinfo {volume} {93}},\ \bibinfo {pages} {021902}
  (\bibinfo {year} {2016})}\BibitemShut {NoStop}%
\bibitem [{\citenamefont {Stephanov}\ and\ \citenamefont
  {Yin}(2018)}]{Stephanov2018}%
  \BibitemOpen
  \bibfield  {author} {\bibinfo {author} {\bibfnamefont {M.}~\bibnamefont
  {Stephanov}}\ and\ \bibinfo {author} {\bibfnamefont {Y.}~\bibnamefont
  {Yin}},\ }\href {\doibase 10.1103/PhysRevD.98.036006} {\bibfield  {journal}
  {\bibinfo  {journal} {Phys. Rev. D}\ }\textbf {\bibinfo {volume} {98}},\
  \bibinfo {pages} {036006} (\bibinfo {year} {2018})}\BibitemShut {NoStop}%
\bibitem [{\citenamefont {Nahrgang}\ \emph {et~al.}(2019)\citenamefont
  {Nahrgang}, \citenamefont {Bluhm}, \citenamefont {Sch\"afer},\ and\
  \citenamefont {Bass}}]{Nahrgang2019hf}%
  \BibitemOpen
  \bibfield  {author} {\bibinfo {author} {\bibfnamefont {M.}~\bibnamefont
  {Nahrgang}}, \bibinfo {author} {\bibfnamefont {M.}~\bibnamefont {Bluhm}},
  \bibinfo {author} {\bibfnamefont {T.}~\bibnamefont {Sch\"afer}}, \ and\
  \bibinfo {author} {\bibfnamefont {S.~A.}\ \bibnamefont {Bass}},\ }\href
  {\doibase 10.1103/PhysRevD.99.116015} {\bibfield  {journal} {\bibinfo
  {journal} {Phys. Rev. D}\ }\textbf {\bibinfo {volume} {99}},\ \bibinfo
  {pages} {116015} (\bibinfo {year} {2019})}\BibitemShut {NoStop}%
\bibitem [{\citenamefont {Shuryak}\ and\ \citenamefont
  {Torres-Rincon}(2019)}]{Shuryak2019prc}%
  \BibitemOpen
  \bibfield  {author} {\bibinfo {author} {\bibfnamefont {E.}~\bibnamefont
  {Shuryak}}\ and\ \bibinfo {author} {\bibfnamefont {J.~M.}\ \bibnamefont
  {Torres-Rincon}},\ }\href {\doibase 10.1103/PhysRevC.100.024903} {\bibfield
  {journal} {\bibinfo  {journal} {Phys. Rev. C}\ }\textbf {\bibinfo {volume}
  {100}},\ \bibinfo {pages} {024903} (\bibinfo {year} {2019})}\BibitemShut
  {NoStop}%
\bibitem [{\citenamefont {Rajagopal}\ \emph {et~al.}(2020)\citenamefont
  {Rajagopal}, \citenamefont {Ridgway}, \citenamefont {Weller},\ and\
  \citenamefont {Yin}}]{Rajagopal2020}%
  \BibitemOpen
  \bibfield  {author} {\bibinfo {author} {\bibfnamefont {K.}~\bibnamefont
  {Rajagopal}}, \bibinfo {author} {\bibfnamefont {G.~W.}\ \bibnamefont
  {Ridgway}}, \bibinfo {author} {\bibfnamefont {R.}~\bibnamefont {Weller}}, \
  and\ \bibinfo {author} {\bibfnamefont {Y.}~\bibnamefont {Yin}},\ }\href
  {\doibase 10.1103/PhysRevD.102.094025} {\bibfield  {journal} {\bibinfo
  {journal} {Phys. Rev. D}\ }\textbf {\bibinfo {volume} {102}},\ \bibinfo
  {pages} {094025} (\bibinfo {year} {2020})}\BibitemShut {NoStop}%
\bibitem [{\citenamefont {Bluhm}\ \emph {et~al.}(2020)\citenamefont {Bluhm}
  \emph {et~al.}}]{Bluhm:2020mpc}%
  \BibitemOpen
  \bibfield  {author} {\bibinfo {author} {\bibfnamefont {M.}~\bibnamefont
  {Bluhm}}  {\it et~al.},\ }\href {\doibase 10.1016/j.nuclphysa.2020.122016}
  {\bibfield  {journal} {\bibinfo  {journal} {Nucl. Phys.}\ }\textbf
  {\bibinfo {volume} {A1003}},\ \bibinfo {pages} {122016} (\bibinfo {year}
  {2020})}\BibitemShut {NoStop}%
\bibitem [{\citenamefont {Du}\ \emph {et~al.}(2020)\citenamefont {Du},
  \citenamefont {Heinz}, \citenamefont {Rajagopal},\ and\ \citenamefont
  {Yin}}]{Du2021}%
  \BibitemOpen
  \bibfield  {author} {\bibinfo {author} {\bibfnamefont {L.}~\bibnamefont
  {Du}}, \bibinfo {author} {\bibfnamefont {U.}~\bibnamefont {Heinz}}, \bibinfo
  {author} {\bibfnamefont {K.}~\bibnamefont {Rajagopal}}, \ and\ \bibinfo
  {author} {\bibfnamefont {Y.}~\bibnamefont {Yin}},\ }\href {\doibase
  10.1103/PhysRevC.102.054911} {\bibfield  {journal} {\bibinfo  {journal}
  {Phys. Rev. C}\ }\textbf {\bibinfo {volume} {102}},\ \bibinfo {pages}
  {054911} (\bibinfo {year} {2020})}\BibitemShut {NoStop}%
\end{thebibliography}

%

\end{document}